\documentclass[11pt,dvips]{article}
\usepackage{amssymb,amsmath}
\usepackage{latexsym}
\newcommand{\beg}[2]{\begin{equation}\label{#1}#2\end{equation}}
\newcommand{\rref}[1]{(\ref{#1})}

\author{Igor Kriz\footnote{Supported by NSF grant DMS 0305853; 
E-mail:ikriz@umich.edu}\\
\small
Department of Mathematics\\
\small University of Michigan\\
\small 2074 East Hall,
530 Church Street\\
\small Ann Arbor, MI 48109-1043,U.S.A.,
\vspace{3mm} \\
\small Michigan Center for Theoretical Physics\\
\small 3444B Randall Laboratory\\
\small University of Michigan\\
\small 450 Church Street\\
\small Ann Arbor, MI 48109-1040, U.S.A. 
\normalsize
\vspace{3mm}
\\
and 
\vspace{3mm}
\\
Hao Xing\footnote{E-mail:haoxing@umich.edu}\\
\small
Department of Mathematics\\
\small
University of Michigan
\\
\small
2074 East Hall,
530 Church Street\\
\small Ann Arbor, MI 48109-1043,U.S.A.
\normalsize
}
\title{On effective F-theory action in type IIA compactifications}

\newcommand{\R}{\mathbb R}
\newcommand{\C}{\mathbb C}
\newcommand{\Q}{\mathbb Q}
\newcommand{\Z}{\mathbb Z}
\def\r{\rightarrow}

\newcommand{\cform}[3]{\begin{array}{c}
{\scriptstyle #3}\\
#1\\
{\scriptstyle #2}\end{array}}

\begin{document}

\maketitle

\begin{abstract}
Diaconescu, Moore and Witten proved that the partition function of
type IIA string theory coincides (to the extent checked)
with the partition function of M-theory. One of us (Kriz) and Sati proposed 
in a previous paper a refinement of the IIA partition function 
using elliptic cohomology and
conjectured that it coincides with a partition function 
coming from F-theory.
In this paper, we 
define the geometric
term of the F-theoretical effective action on type IIA compactifications.
In the special case when the first Pontrjagin class of spacetime
vanishes, we also prove a version of the Kriz-Sati conjecture
by extending the arguments of Diaconescu-Moore-Witten. We also
briefly discuss why even this special case allows interesting examples.
\end{abstract}

\section{Introduction}

The purpose of this paper is to carry out the first stage of a program proposed
in \cite{KSmod} of relating F-theory to type II partition
functions. In the groundbreaking paper
\cite{DMW}, Diaconescu, Moore and Witten established a solid connection
between type IIA string theory and M-theory by comparing their partition
functions. The subtlety lies mainly in the phase, which is
determined by topological terms.
In M-theory, we have the Chern-Simons term, in type IIA
we have flux quantization using $K$-theory field
strength. These definitions are quite different, and
yet the partition functions coincide in the range examined, providing concrete
computational evidence for string duality between type IIA and M-theory.

\vspace{3mm}
Yet \cite{DMW} also contains a puzzle: in type IIB, when in the presence
of the $H_3$ NS-NS field, the $K$-theoretical field strength
is given by twisted $K$-theory, and ends up being incompatible with S-duality,
in which $F_3$ and $H_3$ form a doublet. This puzzle was examined in
\cite{KSB} where it was determined that it cannot be solved by merely
modifying the definition of twisted $K$-theory: one needs to quantize
using a genuinely
new generalized cohomology theory. A candidate emerged essentially simultaneously
in the form of elliptic cohomology: in \cite{KS}, the first author
and Sati noticed that the quadratic structure of the
$K$-theory IIA partition function has an analogue in elliptic
cohomology. A similar construction works for IIB also, and has the
intriguing aspect that, at least when we use the theory $TMF$ of topological
modular forms, which can be regarded as a `universal form of elliptic cohomology'
(although it is not really elliptic cohomology in the proper sense, see comments
in Section \ref{st} below),
turning on the $H_3$-field does not result in a twisting, but merely in
multiplication by a cohomology class, which makes it possible that
$H_3$ and $F_3$ could form a doublet in the same theory.

\vspace{3mm}
In \cite{KSmod}, another connection was proposed, with a $12$-dimensional
theory known as F-theory. This theory was first considered by Vafa
\cite{Vafa}. It was proposed that F-theory compactified on an elliptic
curve $E$ is type IIB string theory, and that S-duality results in 
modular transformations on the space of the elliptic curves $E$.
An action for F-theory relevant to that context was proposed in \cite{Sag}.
Further dualities of F-theory compactified on
an elliptic curve with type I and heterotic string theory were
considered in \cite{Sen, FMW}. 

\vspace{3mm}
Based on this evidence, \cite{KSmod} proposed essentially the following program:
The elliptic cohomology refinement of the quadratic structure introduced in
\cite{DMW} leads to a refinement of the $K$-theoretical field strength in
type IIB string theory, and its partition function, which coincides with
(the relevant part of) the partition function of F-theory, which would
confirm the conjectured duality between type IIB and F-theory, simultaneously
solving the S-theory puzzle.

\vspace{3mm}
Yet, this proposal has severe tests to withstand: By T-duality, there
must be some analogous story for type IIA. In fact, there we are better
equipped to study the partition functions, because of the work done in
\cite{DMW}. However, the problem is that one needs a universal definition
of F-theory which would not be restricted to an elliptic curve fibration on
IIB. In particular, we need to explain how the $G_5$ field, which seems
intrinsic in the action of \cite{Sag, KSmod} arises in a version of F-theory
which contains type IIA string theory.

\vspace{3mm}
It is this part of the proposal which is investigated in the present paper.
In other words, one of the main results of this paper is establishing
a connection of F-theory with type IIA.
Starting with an $11$-dimensional M-theory spacetime $Y$, and its 
$12$-dimensional
spin
cobordism $Z$, as in \cite{DMW}, we investigate an analogue of Witten's
index formula \cite{flux} for the corrected Chern-Simons term on the
{\em loop space} $LZ$, i.e. the space of maps from $S^1$ to $Z$. We use
Witten's theory of the Dirac operator on loop space \cite{W} to propose 
an F-theoretical analog of the Chern-Simons term. This is no longer
a topological term (at least not in general), but it is a geometrical
term in the sense that it depends only on the metric on the world
volume. We propose a concrete formula for the
geometric action term of F-theory compactified on type IIA spacetime.
Further, we interpret
the field strength $G_4$ on the loop space $LZ$ as an object
encoding simulatneously the field strengths $G_4$ and
$G_5$ on $Z$, thus establishing the connection with F-theory containing
type IIB. 

\vspace{3mm}
We do in fact investigate the compatibility of a fundamental F-theory
with the basic T-duality between type IIA and IIB string theories.
Thinking of ``IIA'' F-theory spacetime as a manifold with boundary
which is the spacetime of M-theory on $X^{9}
\times S^1 \times S^1$, then we describe a process in which
the first copy of $S^1$ shrinks to a point, while preserving the
bulk. In a suitable case, this process gives F-theory fibered on IIB,
which is one generalization of T-duality. 
We also predict that the resulting theory without
boundary should have a self-T-duality, which would exchange $G_4$ and $G_5$.

\vspace{3mm}
We then define a certain version of the elliptic partition function in type IIA
proposed in \cite{KS}, and indeed show that this partition function
coincides with the partition function of the version of F-theory just
discussed in the special case when the first
Pontrjagin class vanishes. In the case of this rather special
assumption we reduce the discussion to an easy extension
of the methods of Diaconescu-Moore-Witten \cite{DMW}.
We also carry out a brief discussion showing that even
this restrictive case includes interesting examples.

\vspace{3mm}
Our computation is
restricted to certain values of the elliptic moduli parameter
at which the phase factor in question is intrinsically defined
in IIA and does not depend on the
choice of $Z^{12}$.

\vspace{3mm}
The present paper is organized as follows: In section \ref{s2}, we review
Witten's formula for the Chern-Simons term in M-theory, and carry out
a direct computation of the index involved. The direct index
calculation turns out to be quite
helpful in understanding the loop space case, which is treated in
section \ref{s3}. In section \ref{efund}, we discuss
fundamental F-theory and T-duality.
In section \ref{sw}, we review all the evidence we
got so far for elliptic cohomology field strength in type II string theory,
which comes from at least four different directions. In section \ref{st},
we define the version of elliptic cohomology-based partition function
on type IIA which we need here. In section \ref{sf}, we compare it to
the partition function of F-theory defined in section \ref{s3} in the
special case $\lambda=0$. In the Appendix \ref{s2a},
we also discuss the
$\lambda=0$ restriction and show that despite its rather restrictive
nature, it allows non-trivial examples.

\section{A recapitulation of Witten's formula}
\label{s2}

In this section, we shall recall Witten's formula 
\cite{flux, DMW} for the topological
(Chern-Simons) term in the action of $M$-theory,
along with the specific numbers involved. This will
be useful in the next section, where numbers will be
replaced by modular forms.
Let $Y$ be an $11$-dimensional spin manifold with a distinguished
$4$-dimensional class $G$. Then the vanishing 
$$MSpin_{11}K(Z,4)=0$$
implies that there exists a $12$-dimensional spin manifold $Z$ whose
boundary is $Y$. Witten's formula for the Chern-Simons term in M-theory
is
\beg{ecs}{L_{CS}=\frac{1}{6}\int_{Z}G(G^2-\frac{1}{8}(p_{2}(Z)-\frac{1}{4}p_{1}(Z)^{2})
).}

\vspace{3mm}
Consider, then, a $12$-dimensional spin manifold $Z$.
Let $G\in H^4(Z,\Z)$, and let $V$ be the adjoint complex $E_8$-bundle
associated with $G$. Let $I_V$ be the index of the Dirac operator
on $Z$, twisted by $V$. Let, also, $I_{RS}$ be the Rarita-Schwinger
index, which in this case is the index of the Dirac operator on $Z$ twisted by
the complexified tangent bundle minus $4$ copies of the complex trivial
($1$-dimensional bundle).
In this section, we recall the derivation,
by direct computation, of following formula \cite{DMW, flux}:
\beg{ew1}{\frac{1}{6}\int_{Z}G(G^2-\frac{1}{8}(p_{2}(Z)-\frac{1}{4}p_{1}(Z)^{2})
)=\frac{I_{V}}{2}+
\frac{I_{RS}}{4}.
}
The reason for doing so is that 
in order to proceed with an analogue on loop
space, we need a good understanding of this formula.
One should mention that the significance of the formula \rref{ew1} is
that it is basically the motivation for setting \rref{ecs}; naively,
one would only the
leading cubic term in $G$. The correction term, to match the
index expression, was (at least in the context of M-theory) 
derived by Ho\v{r}ava and Witten 
\cite{hw1,hw2} by considering anomaly cancellation in
heterotic M-theory. The other term is a correction term
which is needed to relate the formula to \rref{ew1}. This is needed
to show the independence $\mod \Z$ of \rref{ecs} of the choice of the cobordism
$Z$ (actually, more precisely, one only uses \rref{ecs} as a phase
factor in the quantum partition function, and to this end one would need
to know that \rref{ew1} is always an integer. This is not true, but
one multiplies the phase by the Pfaffian of the Rarita-Schwinger operator,
which makes the phase always $1$; see \cite{DMW}).

\vspace{3mm}
To recapitulate the derivation of
\rref{ew1}, recall that the Atiyah-Singer index theorem
implies that
\beg{ew2}{I_V=\int_{Z}ch(V)\hat{A}(Z).
}
Thus, we are essentially done if we can calculate $ch(V)$. Let us
recall how this is done. The essential point is that this must be
a polynomial in $G$ with integral coefficients. To discover this polynomial,
it is not important that $Z$ be a $12$-manifold,
and in fact it is sufficient to consider the case $Z=BSU(2)$, 
and to take
a map 
$$BSU(2)\r BE_8$$
induced by a homomorphism $\phi:SU(2)\r E_8$ which is a
canonical inclusion induced by a coroot. The map $\phi$ then induces
an isomorphism on $\pi_3$, hence $B\phi:BSU(2)\r BE_8$ induces
isomorphism of $\pi_4$, and hence
on $H^4(?,\Z)$ by the Hurewicz theorem. 

\vspace{3mm}
But then we can further pull back via the map $BS^1\r BSU(2)$ which
is the classifying map of the group homomorphism $S^1\r SU(2)$
given by the maximal torus. 
Therefore, if we pull back to $S^1$, we have an inclusion of $S^1$
to $E_8$ as a direct summand of the maximal torus (the choice of
direct summand doesn't affect the answer).
We need to calculate the 
Chern character of this representation
of $S^1$. This means we must calculate how $S^1$ acts on the Lie algebra $e_8$.
This can be figured out as usual by taking a coroot $h$, and taking the sum
\beg{ew3}{\sum_{\alpha}z^{\langle h,\alpha\rangle}+rank(G)}
where here $G=E_8$ and the
sum is taken over all roots. Taking the standard presentation of the $E(8)$ lattice
as generated by integral vectors $(a_1,...,a_8)$ with $\sum a_i$ even
(the $D(8)$ lattice) and $(1/2,...,1/2)$, we may take the (co)root $h$
to be $(1,1,0,...,0)$, so there is one root whose inner product with $h$ is $2$,
$2\cdot 2\cdot 6$ roots in the $D(8)$ lattice and $2^5$ roots with fractional coordinates
(a total of $56$ roots) whose inner product with $h$ is $1$, an equal number
of roots with opposite signs, and all remaining $126$ roots having $0$ inner
product with $h$. Thus, \rref{ew3} is
\beg{ew4}{z^2 +z^{-2} +56z + 56 z^{-1} +134,
}
and the corresponding Chern character is calculated by plugging in 
$z=e^{G/2}$, i.e. we get
\beg{ew5}{e^G + e^{-G} + 56e^{G/2} +56e^{-G/2} +134.
}
Expanding, \rref{ew5} is equal to
\beg{ew6}{248 + 60G + 6G^2 +\frac{1}{3}G^3 + HOT.
}
Now regarding the $\hat{A}$-class, one has
\beg{ew7}{{
\begin{array}{l}
\hat{A}(Z)=1-\frac{1}{24}p_1(Z) + \\
\frac{1}{5760}(-4p_2(Z) +7p_{1}(Z)^{2}) -\\
\frac{1}{967680}(16p_3(Z) - 44p_1(Z)p_2(Z) + 31 p_1(Z)^{3}) + HOT.
\end{array}}
}
Multiplying with \rref{ew5} and restricting
attention to elements of dimension $12$, we obtain
\beg{ew8}{\frac{I_{V}}{2}=\int_{Z}(\frac{1}{6}G^3-\frac{1}{48}(p_2-
p_{1}^{2}/4)G-\frac{31}{15120}
p_3+ \frac{13}{30240}p_1p_2-\frac{1}{15120}p_{1}^{3}).
}
The Chern character of $TZ_{\C}-4$ is
$$8+p_1+\frac{1}{12}(p_{1}^{2}-2p_2)+\frac{1}{360}(p_{1}^{3}-3p_1p_2+3p_3).
$$
We get
\beg{ew9}{\frac{I_{RS}}{4}=\int_Z\frac{31}{15120}p_3-\frac{13}{30240}p_1p_2+
\frac{1}{15120}p_{1}^{3},
}
so adding \rref{ew8} and \rref{ew9} gives \rref{ew1}.

\section{Lifting the action from M-theory to F-theory}
\label{s3}

The above discussion suggests that to obtain a natural version of the 
action \rref{ew1} which involves a modular parameter of an elliptic curve
$\tau$, we should replace the $\hat{A}$-class in \rref{ew2} by the
Witten class with generating series
\beg{ef1}{\frac{u/2}{sinh(u/2)}\cform{\prod}{n=1}{\infty}
\frac{(1-q^n)^2}{(1-q^ne^u)(1-q^ne^{-u})}=exp\left(\cform{\sum}{k>0}{}\frac{2}{
(2k)!}S_{2k}
u^{2k}
\right).
}
Here we follow Zagier \cite{zag}. We put, as usual, $q=e^{2\pi i\tau}$, but
to avoid confusion, here we replace the notation $G_{k}$ for Eisenstein
series by $S_k$, so we have
\beg{ef2e}{S_{\ell}=S_{\ell}(\tau)=-\frac{B_{\ell}}{2{\ell}} + 
\cform{\sum}{n=1}{\infty}\left(\cform{\sum}{d|n}{}d^{\ell-1}
\right)q^n.
}
In discussing the geometric term of the F-theory effective action,
we will first assume
\beg{ew0}{p_1(Z)=0.} 
This is a technical assumption, under which
there exists the index on untwisted
Dirac operator on loop space, see \cite{W}. 
However, we shall remove the assumption later in this section,
and derive the effective action in the general case, without
any restriction on $p_1(Z)$.
Now we would like
an analogue of the action \rref{ew1}, but we realize immediately
that there is a problem. What should one replace the $I_{RS}$ term
by? Clearly, it is wrong to modify the Rarita-Schwinger index
formula simply by replacing the $\hat{A}$ genus by the Witten genus.
The reason is that the resulting formula would no longer compute
an index on loop space (and consequently, for example, would not
be modular). Instead, we should somehow involve the tangent bundle
of the manifold $LZ$, but we do not know what the right
analogue of the Rarita-Schwinger
index is in that case, since $LZ$ is infinite-dimensional. Clearly,
some additional idea is needed to solve this
problem; we return to this point briefly
at the end of
this section. For the purposes of the present paper, we choose a somewhat
artificial (although, again, suggestive) solution.
Recalling \rref{ew9}, we see that in the finite-dimensional case,
the Rarita-Schwinger index serves to simply cancel the $12$-dimensional
term of the gravitational anomaly. In the loop case, we
encounter a similar term. In the absence of knowing the exact form of the
loop space index which would cancel that term, we simply take only
the contribution of the $V$-bundle on $LZ$, and truncate
the resulting 
loop index formula below the $12$-dimensional term, assuming that we 
have a ``Rarita-Schwinger''-type
index on loop space which would cancel it. This is not so unreasonable
for our purposes, since 
we are attempting to carry out an analysis analogous to \cite{DMW}, where
the main focus is also on the $E_8$-bundle contribution.
Then, we are only interested
in the characteristic class associated with \rref{ew1} up to $p_2$, so we
get
\beg{ef2}{1-\frac{1}{6}S_4p_2.
}
Thus, the modification of the action \rref{ew1} in the presence of
the restriction \rref{ew0} is
\beg{ef3}{\int_{Z}G(\frac{1}{6}G^2-5S_4p_2).
}
(When $q=0$, $S_4=-B_4/8=1/240$, so \rref{ef3} reduces to \rref{ecs}.)
Now let us explain the physical motivation for choosing this formula.
The main idea is that, denoting by $W$ the Witten class, then
\beg{ef4}{\int_{Z}ch(V)W(Z)
}
is supposed to be equal to the index of the Dirac operator on the loop
space $LZ$ (consisting of smooth maps from $S^1$ to $Z$),
twisted by the bundle $V$, pulled back to $LZ$. 
At present, this cannot be made mathematically rigorous, since one cannot
rigorously define the index on loop space, but Segal \cite{sell} explains that
one may conjecture an analog of the localization formula for equivariant index
applies (Proposition 3.8 of \cite{sell}), and that this formula gives
\rref{ef4}. In fact, this presents another small problem, since the 
obvious projection
$LZ\r Z$ is by evaluation at a point, which is not equivariant, so one must
argue why the pullback of the bundle $V$ to $LZ$ can be taken to be an
$S^1$-equivariant bundle. However, one way to treat this is to note that the index
of the Dirac operator is actually determined by
localizing near fixed loops, where the problem
does not arise.

\vspace{3mm}
Let us now turn to the question as to what the action should
look like in the absence of the restriction \rref{ew0}. It is a
standard procedure to try to cancel an anomaly by adding a gauge term.
In the case of the Witten genus, as described in \cite{W},
the anomaly of the Dirac operator on $LZ$ can be cancelled by
subtracting the bundle of loops on a ``gauge bundle'' $V$ on
$Z$ which satisfies the condition
\beg{e*cond}{p_1(V)=p_1(Z).}
However, which bundle to choose? A spin bundle is, of course,
not characterized by its first Pontrjagin class. Additionally,
the obvious candidate suggested by the term ``gauge bundle'',
namely the adjoint vector bundle of an $E_8$-bundle associated
with a $4$-dimensional integral cohomology class, turns out
to give the wrong answer (as can be checked using the 
Pontrjagin class calculation reviewed in the last section).
It is worth commenting here that the term we are seeking
is in fact not the usual string gauge term. What we are trying
to do is cancel the anomaly associated specifically with lifting
the action to the loop space, but 
the whole term we are studying here corresponds to the gauge
term in M-theory action. In other words, we are trying to cancel
an anomaly which, in this lift, arises within the M-theory gauge
term itself.

\vspace{3mm}
Nevertheless, we can actually deduce what Pontrjagin classes
the bundle $V$ must have. The general form of the F-theory action compactified
to IIA spacetime must be
\beg{eff+}{\int_{Z}G(\frac{1}{6}G^2-5S_4(p_2-\frac{1}{4}p_{1}^{2})).
}
To see this, disregarding the dimension of $G$, terms of
topological dimension $2n$ must be coupled to modular
forms of weight $n$; therefore, terms of topological dimension
$4$ are excluded, as there are no modular forms of weight $2$.
Alternately, any such term is a scalar multiple of $p_1$, which is
excluded by the requirement of anomaly cancellation of the Dirac
operator on loop space; the $8$-dimensional term is identified
by taking the $q\r 0$ limit, and is unique, as again there are
no other modular forms of weight $4$. The $12$-dimensional term
is excluded also by the $q\r 0$ limit, as there is no cusp form
(modular form vanishing at $q=0$) of weight $6$. This last argument
also confirms our Ansatz that the loop Rarita-Schwinger operator,
whatever it is, must cancel the 12-dimensional term.

\vspace{3mm}
In other words, the action \rref{eff+} is fixed. If indeed we
were looking for a gauge bundle $V$ accomplishing this transformation,
it would have to have
$$p_1(V)=p_1(Z),\; p_2(V)=\frac{1}{4}p_{1}(Z)^2.$$
We do not know if if such a bundle indeed exists in general, although
rationally (when $p_1(Z)$ is divisible by a certain integer), this
is always possible by general homotopy-theoretical arguments, as
$BSpin$ is rationally the product
$$\cform{\prod}{n>0}{}K(4n,\Q).$$

\vspace{3mm}
Why the index on loop space be related to F theory can be justified in
the following way: thinking of $G$ as a field
strength, the formula \rref{ew1} can be thought of as a part of the action describing
the dynamics of a $3$-dimensional world volume in $Z$. Now considering
the motion of the same world volume $A$ in $LZ$ is the same thing as considering
the motion of $S^1\times A$ in $Z$. This means that if we believe the index
formula on loop space, then \rref{ef3} describes the dynamics of a 
$5$-dimensional field strength $G_5$ in $Z$ (more exactly
its sector where the ``time coordinate'' of the world volume
together with the additional $S^1$-coordinate form an elliptic
curve of modulus $\tau$, similarly as when
calculating the partition function of a fundamental
string). This is a field one expects to
see in F-theory. Denoting the corresponding potential by $A_4$,
this was noted in \cite{KSmod}, following \cite{Sag}.
It was further remarked in \cite{KSmod} that the reason the $G_5$-field appears
is that the F-theory action provides a framework for unification of
IIA and IIB string theories. Similarly as in \cite{W}, the geometric
presence of the elliptic curve 
of modulus $\tau$ in the dynamics of
$G_5$ is the physical cause of the modular invariance of the formula
\rref{ef3}.

\vspace{3mm}
There is, however, a surprise contained in this analysis. In this picture,
we do not see F theory spacetime as fibered over M theory. Rather, the
connection is that if Y is a spin manifold
interpreted as M theory spacetime, then the phase of its
M-theory action (the Chern-Simons term) is computed by integrating
\rref{ew1} over a manifold $Z$ which is a spin cobordism between $Y$ and 
$\emptyset$. The modular formula \rref{ef3} describes the action of
a field which is ``$S^1$-fibered over $G_4$'' in the above sense, but which
is present on $Z$. Therefore, we are not seeing a $12$-dimensional
manifold $Y\times S^1$, and carrying this further, we certainly are not
seeing a $13$-dimensional manifold whose boundary would be a $12$-dimensional
manifold in the present picture.

\vspace{3mm}
One more comment is in order about the formula \rref{ef3}. Recall again that we
didn't discuss the Rarita-Schwinger index. Indeed, it is not obvious
what kind of analogue of the
Rarita-Schwinger operator one should consider on loop space. Observing
\rref{ew8} and \rref{ew9}, however, suggests that we could solve this
problem here simply by dropping the $12$-dimensional term (not
counting the dimension of $G$). 
Perhaps this points, however, to the deeper point that boundary
phenomena become more complicated on loop space, since the ``boundary'' of
the space of loops on a manifold cannot be identified simply with the
loop space on the boundary. Even more fundamentally, the ultimate
problem may be that we do not understand well gravity in F-theory.
Although a possible supergravity was proposed for physical
F-theory in \cite{FK}, it is fair to say that our understanding
of this situation is indeed in its infancy.

\vspace{3mm}

\section{On fundamental F-theory and duality}
\label{efund}

In this section, let us look briefly at what our findings signify 
for the development of F-theory as a fundamental physical theory.
In \cite{Sag}, Ferrara, Minasian and Sagnotti investigated the 
$F$-theory geometric coupling term
\beg{efms1}{\int_{Z^{12}}A_{4}\wedge G_4\wedge G_4
}
where $A_4$ is the $4$-potential of type IIB
lifted to $12$ dimensions, and $G_4$ is the $4$-form of $11$-dimensional
supergravity lifted to $12$ dimensions (i.e. the field we considered above).
Without going into details of their
notation, let us note that they investigate subharmonic expansions of $A_3$, $A_4$
on $Z^{12}=M^6\times CY$ of particular form, and show that \rref{efms1}
then gives an expected $6$-dimensional coupling term, thus justifying
that \rref{efms1} must be present. They also note, by similar arguments, 
that correction terms
of the forms
\beg{efms2}{\int_{Z^{12}} A_4\wedge I_8,
}
\beg{efms3}{\int_{Z^{12}}G_4\wedge I_8,
}
\beg{efms4}{\int_{Z^{12}}G_4\wedge G_4\wedge G_4
}
must be present with suitable coefficients.

\vspace{3mm}
Now one puzzling part of this picture is the relationship between 
the fields $G_4$ and $A_4$: \cite{Sag} note that the fields $A_4$, $G_4$
are not independent in their
example, yet they do not derive an equation relating them.
They propose that such equation should be supplied by the dynamics of
F-theory. Yet, no such equation has emerged since. 

\vspace{3mm}
One of the points of the present paper is to add evidence to the
proposal that there exists a {\em fundamental} physical F-theory. 
This proposal was made in \cite{FK}, as a natural conclusion
following the investigation done in \cite{KS,KSB,KSmod}.
The idea is that theories in $12$ dimension should be somehow unified
to the same theory. In the present paper, we are investigating the
manifold $Z^{12}$ which is a cobordism of a spacetime of M-theory.
This is, of course, not the same thing as the elliptic fibration
on IIB space considered in 
\cite{Sag}. Yet, we propose that they are sectors of the same theory
(later in this section, we shall make that more explicit). 

\vspace{3mm}
For now, let us note that the field content we observed 
in $Z^{12}$ is analogous to the field content of \cite{Sag}:
we obtained $G_5$ as $G_4$ moving on the loop space of
$Z^{12}$. We can visualise this as a $3$-dimensional world volume
$M_3$ in $LZ^{12}$. In terms of field strengths, this suggests
that $G_4$, $G_5$ are components of a unified field
\beg{efms5}{G\in H^{4}(LZ^{12},\Z).
}
(In fact, if we really wish to formulate a fundamental theory,
we must note that $E_8$-bundles on $LZ^{12}$ are no longer classified
by $H^4$, so we must also require the additional quantization
condition that \rref{efms5} lifts to an $E_8$-bundle.)

\vspace{3mm}
Now in our setting, the Chern-Simons term of M-theory seems to have a 
natural generalization to fields of the form \rref{efms5},
which can be considered as a modification
of \rref{efms1}.
The topology of this situation can be interpreted as a tie between $G_4$ and $A_4$
($A_4$ being a potential corresponding to the field strength $G_5$),
but not as strong as a coupling equation.
We therefore conclude that the relationship \cite{Sag} need is 
an attribute of examining a particular sector of the theory, just
as the restriction on $G_4$ assumed in \cite{DMW} (that it come from
a cohomology class of $X^{10}$).

\vspace{3mm}
Of course, one important point we must address is relating our setting
to the setting of \cite{Sag}. There are important differences, for example
the fact that we consider $Z^{12}$ to be a spin cobordism of M-theory spacetime
$Y^{11}$, i.e. a manifold with boundary. However, there is a path we can
use to relate our theory to an elliptic curve fibration of IIB, namely
a generalization of T-duality relating IIA and IIB: when IIA is considered
on a space of the form $X^{9}\times S^1$, by shrinking the $S^1$ to a point,
that coordinate disappears, but
a string wrapped around the $S^1$ becomes light, signalling the opening
of another dimension, thus giving the T-dual IIB theory on $X^9\times S^1$.

\vspace{3mm}
Now it is impossible to apply such T-duality naively to M-theory, because 
M-theory lacks fundamental strings; it has however $2$-branes, and one compactifies
M-theory on $S^1\times S^1$ and shrinks this $2$-torus to a point, the
$2$-brane wrapped on the torus becomes light and new dimension opens
up, giving $10$-dimensional IIB-theory. 

\vspace{3mm}
In our settings, the situation is additionally complicated by the fact that
we are considering 
the $K$-theory field strength corrections to type II strings, so we must 
take into account spin structure. Consider type IIA string theory
on $X^{9}\times S^{1}_{NS}$, which is M-theory on 
\beg{efms6}{X^{9}\times S^{1}_{NS}\times S^{1}_{R}.}
Now we know, however, that this is really F-theory on a spin-cobordism 
$Z^{12}$ between
\rref{efms6} and $0$. In the special case \rref{efms6}, however, a special
kind of $Z^{12}$ can be proposed, namely
\beg{efms7a}{X^9\times E^{\prime}\times S^{1}_{R}
}
where $E^{\prime}$ is a spin-cobordism from $S^{1}_{NS}$ to $0$. Then
shrinking the size of $S^{1}_{NS}$ to $0$ in the boundary
(while preserving the bulk) corresponds to gluing a disk to
$E^{\prime}$. Denoting the corresponding closed surface by $E$ (which can
be an arbitrary Riemann surface, in particular an elliptic
curve), we get the corresponding bulk F-theory on
\beg{efms7}{X^9\times E\times S^{1}_{R}.
}
Thus, assuming the T-duality mechanism extends
to this cobordism 
context, we exhibited a ``T-duality'' between the $F$-theory with boundary
$M$-theory and $F$-theory fibered on IIB spacetime in this case.

\vspace{3mm}
Let us also note however that there should be another ``self-T-duality''
of the fibered $F$-theory. Consider in particular the F-theory on
\beg{efms8}{X^{9}\times \cform{\prod}{i=1}{3}S^{1}_{R}
}
which is a special case of \rref{efms7}. Then from the evidence we saw,
this theory should have a $2$-brane $M_2$ and a $3$-brane $M_3$ where
the relationship \rref{efms5} becomes 
$$M_3= M_2\times S^1.$$
In particular, then, $M_2$ can be wrapped on
$$\cform{\prod}{i=2}{3}S^{1}_{R}$$
and $M_3$ on
$$\cform{\prod}{i=1}{3}S^{1}_{R}.$$
If we shrink the radius of the first copy of $S^{1}_{R}$ to $0$, then,
$M_3$ will lose a dimension, but $M_2$ will expand by the new dimension,
and we see than that the system $(M_2,M_3)$ is self-dual.

\vspace{3mm}

\section{The evidence for elliptic cohomology field strength in string theory}
\label{sw}

Very briefly, much of our present nonperturbative knowledge of string theory
came from the study of $D$-branes. Polchinski \cite{pol1} noticed that the
RR-fields in free type II supergravity can be interpreted as $p$-
dimensional differential forms ($p$ odd for type IIA and even for type IIB).
He proposed that in non-perturbative theory, these forms correspond
to submanifolds which are $D$-branes, i.e. boundary sets of open strings. 
The $D$-brane is however considered as a dynamic object with $1$ timelike dimension,
so a $p$-form corresponds to a $p-1$-brane. On the other hand, the forms
are potentials, so a $p$-form corresponds to $p+1$-dimensional field strength
denoted by $F_{p+1}$ or $G_{p+1}$.
The relationship between potential and field strength has not been completely
mathematically clarified in general dimension, although various models have
been proposed which generalize both the exterior differential and the 
connection-curvature relationship. There are other complications to the story,
such as the distinction between electric and magnetic charges in relationship
to Poincare duality, but we needn't discuss these here. We refer the reader
to \cite{pol2} for a survey.

\vspace{3mm}
Witten however noticed that this story needed a correction (see
\cite{Wi1}). $D$-branes carry
vector bundles of Chan-Paton charges, and in fact $D$-brane processes 
correspond to addition of vector bundles, so in other words the charges
of stable $D$-branes should be elements of $K$-theory. This should then
be true also of the field strengths $G_{k}$. It leads to the conclusion
that all field strength sources in type II string theory which have
apparently different dimensions
can be added and expressed in terms of one $K$-theory class $G\in K^0(X)$
for $X$ of type IIA and $G\in K^1(X)$ for $X$ of type IIB.
A free field approximation partition function for type IIA and IIB based on this
is calculated in \cite{DMW}, and takes the form of a theta function.
The main result of \cite{DMW} is showing that this matches
the partition function of M-theory
(with all the subtlety contained in the discussion of phase).

\vspace{3mm}
The first evidence that an additional correction 
is needed was in fact given in the same paper,
\cite{DMW}. There, Diaconescu, Moore and Witten 
noticed that in the presence of the $H_3$-field
in type II string theory, the correct analog of the $K$-theory charge
lies in twisted $K$-theory where the twisting is given by $H_3$.
But this violates S-duality in type IIB: the first k-invariant, or
``Atiyah-Hirzebruch spectral sequence differential'' in twisted
$K$-theory is not invariant with respect to the modular transformations
between $H_3$ and $F_3$ which $S$-duality predicts. In fact, in
\cite{KSB}, the the first author and Sati showed that there exists no
modified definition of twisted $K$-theory which would remedy this problem
without introducing other homotopy groups (i.e. other fields).

\vspace{3mm}
Why is S-duality violated in the $K$-theory field strength? An explanation
was offered in \cite{KSmod}: the match between the IIA and M-theory
partition functions in 
\cite{DMW} can be interpreted as confirmation of Witten's conjecture that
IIA string theory can be interpreted as compactification of $11$-dimensional M-theory on
a circle. This is a refinement of his earlier observation that IIA theory at
strong coupling
is M-theory. Considering the total field strength in $K$-theory involves
bringing out this non-perturbative view of the theory. But now considering
the same field strength in type IIB suggests (as can be further substantiated
by T-duality) the same construction, i.e. viewing IIB theory as
an $11$-dimensional ``T-dual'' 
of M-theory compactified on a circle. From that point of view,
however, it is apparent why such model should violate S-duality, since
such duality should partner the extra dimension with another: this
is the proposal of Ferrara, Minasian and Sagnotti \cite{Sag} which suggests
to consider IIB instead as a compactification of $12$-dimensional
F-theory on an elliptic
curve whose dimensions are permuted by S-duality.

\vspace{3mm}
Returning however back to type IIB, what kind of refinement of the
$K$-theory field strength could possibly capture the additional $12$'th
dimension, and match the partition function of F-theory? 
In \cite{KSB},
a natural candidate emerged in the form of elliptic
cohomology. Elliptic cohomology (more precisely the
Hopkins-Miller spectrum $TMF$) was suggested from at least four different
directions. First of all, its torsion free classes are modular forms,
so this suggests modularity. However, we must be careful not to identify
this modularity with S-duality in the most obvious way: as we saw above,
the modularity of $TMF$ is a phenomenon in two of the four dimensions of
a world volume, while $S$-duality is a spacetime phenomenon. If
we propose a IIB field strength based in $TMF$, then its topological modularity
will be present even in the analogous field strength of IIB,
which must in fact exist by T-duality.

\vspace{3mm}
More strongly than the modularity, $TMF$ has the property that
turning on the $H_3$-field does not cause twisting, but merely multiplication
by a certain $TMF$ class (see \cite{KSB}). This is encouraging, because
the twisting itself is a puzzle. If we want to turn on both the $H_3$ and
$F_3$ field in type IIB, we want the field strength to be group-valued.
But this is not the case for twisted $K$-theory considered simultaneously
at different twistings: It seems to require a single generalized cohomology
theory which is not twisted by turning on these fields.

\vspace{3mm}
The next piece of evidence for elliptic cohomology came in \cite{hk}, \cite{hk1}
where Hu and the first author attempted a more rigorous mathematical formalism
for $D$-branes. In that paper, it was showed that while a simpler model
of $D$-branes exists with Chan-Paton charges in vector spaces, 
the full model which captures all possible anomalies of $D$-branes
has charges in $2$-vector spaces, the theory of which is 
linked with elliptic cohomology.

\vspace{3mm}
Finally, the first author and Sati \cite{KS, KSB, KSmod}
noticed that an elliptic cohomology-valued field
strength and partition
function can in fact be defined in type IIA and IIB in analogy with
the $K$-theory picture. For technical reasons, actually, these
papers work with real-oriented elliptic cohomology, which has only
level $3$ modularity (i.e. its torsion free classes are
forms defined on the moduli space of elliptic curves
with level $3$ structure. Below we propose a correction in the case of type
IIA based on the theory $TMF$, 
which has full modularity. This partition function matches
(to the extent considered) the
F-theory partition function.

\vspace{3mm}
One should find out if indeed the $TMF$-partition function in IIB 
matches the F-theory partition function, and therefore preserves
S-duality. This problem will be considered in future work. In the
present paper, we specialize to IIA, where the discussion is
easier because of the huge amount of work done in 
\cite{DMW}.

\vspace{3mm}
We should also mention that the original motivation for the elliptic
partition function in \cite{KS} was that the $W_7(X)$ anomaly of
type IIA string theory on $X$ turns out to be precisely the obstruction
to orientability of spacetime $X$ with respect to complex-oriented
elliptic cohomology. However, one needs orientability of $X$
with respect to real-oriented elliptic
cohomology to get a quadratic structure of the kind needed for defining
a theta function. The obstruction to such orientability is $w_4$.
Orientability of $X$ with respect to $TMF$, which we consider here,
has the obstruction $\lambda\mod 24$. These are successively stronger
conditions. The $4$-dimensional obstructions were
previously known to arise in type I and heterotic string theories, but
not in type II. Therefore, this suggests that the F-theory interpretation
should unify all types of string theories. In the present paper, we 
operate with index on loop space, which requires the vanishing of
$\lambda$ outright (the condition \rref{ew0}). This is why we impose
this strongest condition throughout the present paper.

\section{The $TMF$ partition function in type IIA.}
\label{st}

In \cite{KS, KSB, KSmod}, Kriz and Sati proposed analogues of partition functions
of type IIA (and IIB) string theory based on elliptic cohomology. 
A part of the reason however a definitive version could not be proposed was
that it was not clear which elliptic cohomology one should use. 
We remarked that likely the Hopkins-Miller theory $TMF$ (which stands
for topological modular forms) should be used, but couldn't do that in
part because $TMF$ is not an elliptic cohomology theory in the ordinary
sense (rather, it is a ``homotopy inverse limit of all elliptic cohomology
theories'', while a universal elliptic cohomology theory does not exist).

\vspace{3mm}
In the present paper, however, we have proposed a much more concrete formula
for the relevant sector of F-theory action, and also understand its physical
interpretation better. In the present context, we have seen that indeed
the index of Dirac operator on loop space occurs, which by Witten's
conjecture is related to the $TMF$ characteristic class. Therefore,
we conclude that we must indeed use $TMF$ itself.

\vspace{3mm}
Now we are mostly interested in $2$-torsion considerations, for which purpose
we can replace $TMF$ by its $2$-complete version in the sense of homotopy
theory. By results of Hopkins and Mahowald \cite{eo2}, this spectrum (which
they call $EO_2$) can be obtained as a certain fixed point spectrum
$(E_2)^G$ where $E_2$ is Landweber elliptic cohomology with coefficients
$W_2[[u_1]][u,u^{-1}]$, $W_2$ are the Witt vectors, $dim(u)=2$, and 
$G=SL_2(\mathbb{F}_3)$ acts on this spectrum in a suitable way. In \cite{eo2},
the most explicit computation of the $2$-completed homotopy groups of $TMF$
are actually given in Theorem 9.11, which describes its connective version.
One obtains $TMF$ from this description by inverting $v_2^{32}$.

\vspace{3mm}
In order to define a $TMF$-valued $\theta$-function, we need a ``complex''
version $TMF_{\C}$ of $TMF$ in the same sense as complex $K$-theory is
a complex version of real $KO$-theory. It is not completely obvious
how to do this for $TMF$ for the following reason: one has
a standard $\Z/2$-action on $E_2$ which corresponds to taking inverse in
the formal group law, and in \cite{real} we called the corresponding
$\Z/2$-equivariant spectrum $E\R_2$. (It is then natural to call the
fixed point spectrum $EO_2$, which unfortunately conflicts with the notation
of \cite{eo2}.) Now the problem is that the inclusion $\Z/2\subset G$
is simply the center (which of course doesn't split
off), so it is not obvious how, simply by using
the equivariant structure, we could ``forget'' the action of this $\Z/2$-group
while remembering the rest of the $G$-action.

\vspace{3mm}
Fortunately, there is a solution stemming from homotopy theory. Let us
denote by $\eta$ the generator of the stable homotopy
group $\pi_1(S)=\Z/2$, and denote (in agreement with \cite{eo2}), by
$M(\eta)$ the spectrum which is the cofiber of this map.
Now in $K$-theory,
we have the relation
\beg{et1}{KO\wedge M(\eta)\simeq K.
}
Now when we invert the class $(v_1)^4$ in the ring
spectrum $TMF$, forming a spectrum
\beg{et1a}{v_{1}^{-1}TMF,}
we obtain a direct (wedge) sum of copies of orthogonal $K$-theory $KO$.
Our idea is that the spectrum \rref{et1a} will be enough to recover our
partition function: we are using index-theoretic tools, which means that
we are essentially using $K$-theory anyway. 
Therefore, it is natural to simply put 
\beg{et2}{v_{1}^{-1}TMF_{\C}= v_{1}^{-1}TMF\wedge M(\eta).
}
We know that \rref{et2} is a commutative associative
ring spectrum (giving a cohomology theory which, on spaces, has graded-commutative
associative multiplication), since the same is already true for the spectrum
$M(\eta)$.
Now the spectrum $v_{1}^{-1}TMF_{\C}$ is a direct (wedge) sum of copies of
complex $K$-theory $K$ (to see this, one can compute, in a stadard way, using
e.g. the Adams-Novikov spectral sequence,
the homotopy of the function spectrum $F(v_{1}^{-1}TMF_{\C}, K)$.
Another observation is that $v_{1}^{-1}TMF_{\C}$ is a complex oriented
spectrum. Furthermore, for 
\beg{et2a}{u\in v_{1}^{-1}TMF_{\C}^{0}(X),}
we have a class
\beg{et2b}{u\overline{u}\in v_{1}^{-1}TMF^{0}(X).}

\vspace{3mm}
Looking at Theorem 9.11 of \cite{eo2}, we see that in the homotopy of
$TMF$, there is copy of the connective $k$-theory homotopy shifted by
the generator $v_{1}^{4}$ which is the Bott class in dimension $8$ (tensored
with $\Z[v_{2}^{4}, v_{2}^{-4}]$, but we don't use that at the moment).
The point is however that there is the class 
\beg{et3}{v_{1}^{4}\eta^2\in TMF_{10}
} 
analogously as in $KO(10)$. Moreover, in $TMF_{\C}$, the 
$$v_{1}^{4}kO_{*}[v_{2}^{4},
v_{2}^{-4}]$$
summand is replaced by 
$$v_{1}^{4}k_*[v_{2}^{4},v_{2}^{-4}],$$
so there also is the non-torsion class
\beg{et4}{v_{1}^{5}\in (TMF_{\C})_{10}.
}
But this suits us well. Now we can take advantage of the fact that, just
as in \cite{KS}, the $\mod 2$ index can be interpreted as cap product 
with an orientation class 
\beg{et5}{[X]\in TMF_{10}(X)
}
where $X$ is $10$-dimensional spacetime (which is assumed to be a compact
manifold which is $TMF$-orientable by \rref{ew0}, \cite{ahr}).

\vspace{3mm}
In the present paper, we only consider a contribution to the modular
partition
function of type IIA string theory based on a $4$-dimensional cohomology
class
\beg{et6}{G\in H^4(X^{10},\Z),}
analogously to Chapter 7 of \cite{DMW}. Additionally, we agreed to invert
$(v_{1})^{4}$, which should preserve all information which we can
recover by index theory. Now \cite{DMW} start with a lift of $G$ to
$K^0(X)$ via the Atiyah-Hirzebruch spectral sequence. We need to form
the lattice $L$ spanned by all such classes $G$ and their duals $*G$. 
Now we know from the above observations that the classes in $L$ in
fact automatically lift to $v_{1}^{-1}TMF_{\C}$. Now we can consider the
``phase factor'' for 
$$u\in v_{1}^{-1}TMF_{\C}$$
given by
\beg{et7}{j(u)=\langle u\overline{u},[X]_{v_{1}^{-1}TMF}\rangle\in v_{1}^{-1}TMF_{10}
}
(see \rref{et2b}). Since we remarked that
$v_{1}^{-1}TMF$ is a wedge sum of copies of $KO$-theory,
we can also interpret this as a $\mod 2$ index where, at least morally, 
again $\hat{A}$ is
replaced by the Witten class $W$. Conjecturally, this should also equal
the $\mod 2$ index on the loop space of $X$.

\vspace{3mm}
There is a subtle problem with this definition. Coefficients of
$v_{1}^{-1}TMF$ are not exacly modular forms. They are
modular functions, but inverting $v_{1}$ corresponds to inverting $S_4$,
i.e. allowing a
pole at $\tau=\root^{3}\of{-1}$ (which is a zero of $S_4$).
If we want to show that we have an actual modular form without singularities,
we
need to know that on the right hand side
of \rref{et7}, we actually get a $v_{1}$-invertible
class in 
$TMF_{10}$ itself, without a $v_{1}$ denominator (i.e. a $v_1$-integral class).
But to this end, we use the definition in \cite{KS}: 
$u$ also lifts into the complex-oriented generalized cohomology theory
$E_{2}$. This is because $E_2$ is a complex-oriented
cohomology theory. But a $4$-dimensional integral class on
a $10$-manifold cannot
have a differential $d_{>3}$ in the Atiyah-Hirzebruch spectral
sequence ($d_3$ is the obstruction to lifting to $k$-theory). 
The reason $d_4$ is impossible is that it would land in $H^8(X,\Z)$
which has a Poincare coupling with $H^2(X,\Z)$. But classes in
$H^2(X,\Z)$ are represented by maps to $\C P^{\infty}$, and the
fundamental class in $H^2(\C P^{\infty},\Z)$ lifts to any complex-oriented
cohomology theory, which makes it impossible for the dual class in $H^8(X,\Z)$
to be killed by an AHSS differential. The argument for $d_{>4}=0$ is analogous.

\vspace{3mm}
Now in \cite{KS}, we constructed an analogue of the
function \rref{et7} which lies in $(E\R_{2})_{10}$ where $E\R_2$ is the
real version of the elliptic cohomology spectrum
$E_2$. Now the coefficients of $E\R_2$ are not modular, but the following
argument can be made: The coefficients of $v_{1}^{-1}TMF$ map injectively to the
coefficients of $v_{1}^{-1}E\R_2$. Moreover, the intersection of the
image with the image of the coefficients of $E\R_2$ is equal to the image
of the coefficients of $TMF$. To see this, the statement is certainly
true rationally, i.e. when we tensor with $\Q$. To obtain the integral
statement, we argue that the torsion free and $v_{1}$-torsion free factor $\Phi_{1}$
of $TMF_*$ is a direct summand of the torsion free and $v_{1}$-torsion free
summand $\Phi_2$ of $(E\R(2))^{\Z/2}_{*}$. Indeed, the latter graded abelian
group is
\beg{esum1}{(\Z\{1,2v_{1}^{1}\}[v_{1}^{4},v_{2}^{2}]
\oplus v_{2}^{2}\Z\{1,v_{2}^{2}\}[v_{2}^{4}])[v_{2}^{-4}],
}
while the former is
\beg{esum2}{(\Z[v_{2}^{32}]\oplus v_{1}^{4}\Z\{1,2v_{1}^{2}\}[v_{2}^{4}])[v_{2}^{-32}].
}
We see that \rref{esum2} is a direct summand of \rref{esum1}. But this means that
$$\Phi_{1}=\Phi_{2}\cap (\Phi_{1}\otimes\Q).$$
The right hand side is, by our rational observation, the intersection of
$\Phi_{2}$ with $v_{1}^{-1}TMF\otimes \Q$, which contains $\Phi_{2}\cap v_{1}^{-1}TMF$,
which is therefore contained in $\Phi_{1}$. The converse inclusion is trivial.

\vspace{3mm}
Therefore, integrality with respect to $v_1$
of the class \rref{et7} can be tested by looking at $E\R_2$, which was
done in \cite{KS}.

\section{Comparing F-theory and IIA.}
\label{sf}

First, let us see what we have computed. Starting with the lattice $L$,
we can take the pairing whose imaginary part, similarly as in
\cite{DMW}, Section 7.1, is given by
\beg{ec1}{\omega(x,y)=\langle x\overline{y} ,[X]_{v_{1}^{-1}TMF_{\C}}\rangle
}
where on the right hand side, we are really considering the lifts of
$x,\overline{y}$ to $v_{1}^{-1}TMF_{\C}$. We observe, however, that the
right hand side is given by the formula
\beg{ec2}{\int_{X}ch(x\overline{y})W(X).
}
Now since we are assuming \rref{ew0}, the lowest dimension in which $W(X)$
differs from $\hat{A}(X)$ is dimension $8$, while the bundles $x$, $\overline{y}$
which we choose to represent the cohomology classes, have $c_1=0$
(see \cite{DMW}, Section 7.2). Thus, we see that we may in effect replace
$W(X)$ by $\hat{A}(X)$ in \rref{ec2}, and the only new information our
partition function will gain will be in the $\mod 2$ index, i.e. in calculating
$j(x)$.

\vspace{3mm}
Now in the last section, we have defined $j(x)$ as an element of the
image of $TMF_{10}$ in $v_{1}^{-1}TMF$. This means that the answer is a `modular
form' (with possible pole at $q=0$), but it is also $2$-torsion! At first,
this seems like a puzzle both mathematically and physically. Mathematically,
clearly one cannot state modularity the usual way in terms of the modular parameter
$\tau$. Yet, the very existence of $TMF$ and its ``modularity'' shows that some
interpretation must exist. The interpretation, of course, is that the class
constructed is of the form $\eta^2\phi$ where 
$\eta$ is the generator of $\pi_1(S)$ and $\phi$ is a modular form with
integral coefficients: one can consider elliptic curves over $\Z$,
and then one has well defined modular forms with
inverted discriminant (modular functions), which, as recalled for
example in \cite{eo2}, form the ring 
\beg{ec3a}{R[\Delta^{-1}]}
where $R$ is the ring of modular forms, given by
\beg{ec3}{R=\Z[S_4,S_6,\Delta]/(1728\Delta=S_{4}^{3}-S_{6}^{2}).}
Of course, we know that not all of the modular forms \rref{ec3a}, even
after multiplication by $\eta^2$, are topological modular forms.
Therefore, the way we must interpret $j(x)$ in the presence of $TMF$ is
that after plugging integral values into topological modular forms, or, 
more precisely, choosing a homomorphism 
\beg{ec4}{h:TMF_{*}\r\Z/2,}
$j(x)=j_{h}(x)$ becomes a well defined integer $\mod 2$. We then obtain
a partition function with phase
\beg{ec5}{\phi_{IIA}^{h}}
in the theta function
\beg{ec5a}{exp(-i\pi Re\tau(\theta_{h}/2))\cform{\sum}{x\in\Gamma_1}{}
exp(i\pi\tau(x+\theta/2))\Omega_{h}(x)}
indexed by the homomorphisms $h$. Here the formula \rref{ec5a} is
precisely analogous to formula (7.11) of \cite{DMW}, where we pointed
out with the subscript $h$ the terms which depend on that homomorphism. 
Physically, we must conclude that
also, a choice of the discrete homomorphism \rref{ec4} is needed to make the phase
\rref{ec5} well defined. We must therefore conclude that from the IIA point
of view, only a discrete set of choices of modular parameter are allowed, and
that restrictions on the geometry of the elliptic curve allowed in thus
``quantized''. The explanation is that the F-theory action term gotten
as a loop version of the Chern-Simons term is geometrical, but in general
not topological: there is therefore no reason why one should be able to
recover the corresponding partition function for IIA. For special values
of the modular parameter, however, there is more symmetry and we shall see
that a relation does hold.

\vspace{3mm}
Let us now set out to carry out an analog of the comparison \cite{DMW}
between type IIA and M-theory phases for modular type IIA and F-theory
phases. 
Unfortunately, we only know how to do the easiest case, namely
when the condition \rref{ew0} holds. A brief discussion of the significance
of this restriction is given in the Appendix.
Under this restriction,
much of the discussion is directly analogous to Section 7 of 
\cite{DMW}. In fact, the discussion somewhat simplifies due to that
fact that we assumed \rref{ew0}. In any case, we start with a class
\beg{ec6}{G\in H^4(X,\Z)}
which has a $K$-theory lift $x$. We represent, again, $x=E-F$ where
$F$ is a trivial rank $5$ bundle. In analogy with formula (7.28) of
\cite{DMW}, we obtain the formula
\beg{ec7}{\Omega_h(x)=(-1)^{q_L(ad(E))+I(E)}
}
where $q_L$ is the $\mod 2$ loop index and $I(E)$ is ordinary index.
Note that the use of ordinary index is justified by the above dimensional
considerations. Similarly, the F-theoretical phase is, in analogy with
formula (7.29) of \cite{DMW}, 
\beg{ec8}{(-1)^{f_h(a)}=(-1)^{q_{L}(ad(E))+I(\wedge^{2}E)}.
}
Therefore, one can carry out a comparison of phases precisely analogous
to \cite{DMW}, Section 7.7, noting that the formula of all index terms
other than the $\mod 2$ index are the same as there, and that the
terms given by the $\mod 2$ index are already matched by comparison
of the formulas \rref{ec7}, \rref{ec8}. Therefore, we conclude that
for these moduli $\tau$ for which
the F-theory phase is only $\pm 1$, there is indeed a match 
between the phases of the $TMF$-based
type IIA partition function and the F-theory phase.

\section{Appendix: Discussion of the condition $\lambda=0$}

\label{s2a}

Let us discuss in this section the condition \rref{ew0}, when
the ungauged Dirac operator on loop space is anomaly free. 
This is the special case in which we proved the Kriz-Sati
conjecture in the last section. The restriction was not
needed elsewhere in the paper. 

\vspace{3mm}
First, let us comment that the condition is actually more accurately
formulated as follows: in a spin manifold, there is a 4-dimensional
characteristic class $\lambda$ such that $2\lambda=p_1$. The
more precise formulation of the condition \rref{ew0} is 
\beg{ew0a}{\lambda(Z)=0.
}
In mathematics, in the context of elliptic cohomology,
manifolds satisfying \rref{ew0a} became known as {\em string manifolds},
a term which since has been occasionally used in physics as well.

\vspace{3mm}
From the point of view of classical string theory, however, this
is a serious misnomer, since the condition actually excludes many 
interesting vacua, for example complete intersection Calabi-Yau
3-folds: Let us, indeed, look at a complete intersection $K$ of
$n-3$ hypersurfaces in $\C P^n$. If these are given by
constraints $F_1,...,F_{n-3}$ of degrees $\ell_1,...\ell_{n-3}$,
then the total Chern class of $K$ is
\beg{eapp1}{c(K)=(1+a)^{n+1}/(1+\ell_1 a)...(1+\ell_{n-3}a)
}
where $a$ is the generator of $H^2(\C P^n)$.
The condition for Calabi-Yau is $c_1=0$, i.e.
\beg{eapp2}{n+1=\cform{\sum}{i=1}{n-3}\ell_i.
}
(We note here that in dimensions $\leq 6$, the 
restriction map $H^*(\C P^n,\Z)\r H^*(K,\Z)$
is an injection: The fundamental homology class $[K]\in
H_6(K)$ maps to a homology class Poincare dual to 
$c_1(F_1)...c_1(F_{n-3})=a^{n-3}\ell_1...\ell_{n-3}$,
which is non-zero, hence $a^3$ restricts to non-zero.
Hence, so must $a,a^2$ by multiplicativity.)
Now from \rref{eapp1}, $c_2(K)=0$ gives
\beg{eapp3}{
\cform{\sum}{i<j}{}\ell_i \ell_j=\frac{n(n+1)}{2}.
}
So from \rref{eapp2}, \rref{eapp3},
$$\begin{array}{l}(n+1)^2=\cform{\sum}{i=1}{n-3}\ell_{i}^{2}+\cform{\sum}{i<j}{}
2\ell_i
\ell_j= \\
\cform{\sum}{i<j}{}\frac{\ell_{i}^{2}+\ell_{j}^{2}}{n-4}+\cform{\sum}{i<j}{}
2\ell_i\ell_j\geq\\
(\cform{\sum}{i<j}{}2\ell_i\ell_j)\frac{n-3}{n-4}=n(n+1)\frac{n-3}{n-4}.
\end{array}
$$
This gives $n^2-3n\leq n^2-3n-4$, which is a contradiction.

\vspace{3mm}
This is, of course, only a most basic example of a fairly standard
computation in this context. In fact, one may show more strongly
that Calabi-Yau $3$-folds in the strong sense, i.e. $6$-dimensional manifolds 
with a metric which has $SU(3)$-holonomy, are flat (see \cite{kob,wil}).

\vspace{3mm}
On the other hand, it is not difficult to show by general
homotopy-theoretical methods that $6$-dimensional homotopy Calabi-Yau
manifolds (also sometimes referred to as weakly Calabi-Yau,
i.e. complex manifolds with $c_1=0$) satisfying \rref{ew0a} exist:
Let us look at the cobordism ring $MU\langle 6\rangle$ of stably weakly
complex manifolds with $c_1=c_2=0$. Let us consider the situation rationally
(more precise arguments are of course available, but would take longer).
By the Thom isomorphism, one has
\beg{ecob1}{H^*(MU\langle 6\rangle,\Q)= H^*(MU\langle 6\rangle,\Q)=\Q[c_3,c_4,...]}
where $BU\langle 6\rangle$ is the classifying space of complex bundles
satisfying $c_1=c_2=0$. Now $n$-dimensional stably weakly complex manifolds 
(i.e. manifolds with complex structure on the stable normal bundle) 
satisfying $c_1=c_2=0$ are classified by the $n$'th stable homotopy group
\beg{ecob2}{\pi_n MU\langle 6\rangle.}
When tensoring with $\Q$, however, stable homotopy groups coincide with
homology, so \rref{ecob1} shows that there are non-trivial elements, in
fact elements $M$ which project non-trivially to $\pi_n(MU)$. That group is
detected completely by Chern numbers. 

\vspace{3mm}
Therefore, we certainly know that there are complex manifolds $K$ with
$c_1(K)=c_2(K)=0$ and which have non-zero Chern number $[c_3(K)]$, i.e.
are not complex-cobordant to $0$. While this shows that there are
$6$-dimensional Calabi-Yau spaces to which our proof applies, this
is however of little interest since the statement trivializes in this
case.

\vspace{3mm}
However, one can obtain non-trivial examples with $8$-dimensional homotopy
Calabi-Yau manifolds. In this case,
because we assumed the
vanishing of $c_1$, $c_2$, the only possibly non-trivial Chern
number is $c_4[M]$ which shows that we must have
\beg{ecob3}{c_4(M)\neq 0\in H^8(M).}
Recall further that the total Pontrjagin class of a complex manifold
has the form $c(M)\overline{c(M)}$ where $c(M)=1+\sum c_i(M)$ and
$\overline{c(M)}=1+\sum (-1)^i c_i(M)$. In view of vanishing of $c_1$, $c_2$,
we see that in our case $p_2=2c_4$ and thus \rref{ecob3} implies
$p_2\neq 0$, so there are non-trivial $8$-dimensional Calabi-Yau
vacua to which our proof applies.

\vspace{3mm}
Although we proved only the easiest case of the Kriz-Sati
conjecture here, nevertheless this discussion may indicate that we are
ultimately touching on the question of physical vacua of F-theory.
Although that subject is not well understood,
it appears that other vacua than coming from 6-dimensional
Calabi-Yau spaces are interesting in that case (\cite{FK}).


\end{document}